\documentclass[revtex4-1]{emulateapj}

\usepackage{graphicx}
\usepackage{subfigure}
\usepackage[pdftex,backref,breaklinks,colorlinks,citecolor=blue,linkcolor=red]{hyperref}
\usepackage{natbib}
\usepackage{hhline}

\usepackage{textcomp}

\begin{document}

\title{TeV Gamma-ray Observations of The Galactic Center Ridge By VERITAS}
\author{
A.~Archer\altaffilmark{1},
W.~Benbow\altaffilmark{2},
R.~Bird\altaffilmark{3},
M.~Buchovecky\altaffilmark{4},
J.~H.~Buckley\altaffilmark{1},
V.~Bugaev\altaffilmark{1},
K.~Byrum\altaffilmark{5},
J.~V~Cardenzana\altaffilmark{6},
M.~Cerruti\altaffilmark{2},
X.~Chen\altaffilmark{7,8},
L.~Ciupik\altaffilmark{9},
E.~Collins-Hughes\altaffilmark{3},
M.~P.~Connolly\altaffilmark{10},
J.~D.~Eisch\altaffilmark{6},
A.~Falcone\altaffilmark{11},
Q.~Feng\altaffilmark{12},
J.~P.~Finley\altaffilmark{12},
H.~Fleischhack\altaffilmark{8},
A.~Flinders\altaffilmark{13},
L.~Fortson\altaffilmark{14},
A.~Furniss\altaffilmark{15},
G.~H.~Gillanders\altaffilmark{10},
S.~Griffin\altaffilmark{16},
J.~Grube\altaffilmark{9},
G.~Gyuk\altaffilmark{9},
N.~H{\aa}kansson\altaffilmark{7},
D.~Hanna\altaffilmark{16},
J.~Holder\altaffilmark{17},
T.~B.~Humensky\altaffilmark{18},
M.~H{\"u}tten\altaffilmark{8},
C.~A.~Johnson\altaffilmark{19},
P.~Kaaret\altaffilmark{20},
P.~Kar\altaffilmark{13},
N.~Kelley-Hoskins\altaffilmark{8},
M.~Kertzman\altaffilmark{21},
D.~Kieda\altaffilmark{13},
M.~Krause\altaffilmark{8},
F.~Krennrich\altaffilmark{6},
S.~Kumar\altaffilmark{17},
M.~J.~Lang\altaffilmark{10},
S.~McArthur\altaffilmark{12},
A.~McCann\altaffilmark{16},
K.~Meagher\altaffilmark{22},
J.~Millis\altaffilmark{23,23},
P.~Moriarty\altaffilmark{10},
R.~Mukherjee\altaffilmark{24},
D.~Nieto\altaffilmark{18},
R.~A.~Ong\altaffilmark{4},
N.~Park\altaffilmark{25},
V.~Pelassa\altaffilmark{2},
M.~Pohl\altaffilmark{7,8},
A.~Popkow\altaffilmark{4},
E.~Pueschel\altaffilmark{3},
J.~Quinn\altaffilmark{3},
K.~Ragan\altaffilmark{16},
G.~Ratliff\altaffilmark{9},
P.~T.~Reynolds\altaffilmark{26},
G.~T.~Richards\altaffilmark{22},
E.~Roache\altaffilmark{2},
J.~Rousselle\altaffilmark{4},
M.~Santander\altaffilmark{24},
G.~H.~Sembroski\altaffilmark{12},
K.~Shahinyan\altaffilmark{14},
A.~W.~Smith\altaffilmark{27,*},
D.~Staszak\altaffilmark{16},
I.~Telezhinsky\altaffilmark{7,8},
J.~V.~Tucci\altaffilmark{12},
J.~Tyler\altaffilmark{16},
V.~V.~Vassiliev\altaffilmark{4},
S.~P.~Wakely\altaffilmark{25},
O.~M.~Weiner\altaffilmark{18},
A.~Weinstein\altaffilmark{6},
A.~Wilhelm\altaffilmark{7,8},
D.~A.~Williams\altaffilmark{19},
B.~Zitzer\altaffilmark{5}
\linebreak
and
\linebreak
F.~Yusef-Zadeh\altaffilmark{28}
}

\altaffiltext{*}{Corresponding Author: asmith44@umd.edu}
\altaffiltext{1}{Department of Physics, Washington University, St. Louis, MO 63130, USA}
\altaffiltext{2}{Fred Lawrence Whipple Observatory, Harvard-Smithsonian Center for Astrophysics, Amado, AZ 85645, USA}
\altaffiltext{3}{School of Physics, University College Dublin, Belfield, Dublin 4, Ireland}
\altaffiltext{4}{Department of Physics and Astronomy, University of California, Los Angeles, CA 90095, USA}
\altaffiltext{5}{Argonne National Laboratory, 9700 S. Cass Avenue, Argonne, IL 60439, USA}
\altaffiltext{6}{Department of Physics and Astronomy, Iowa State University, Ames, IA 50011, USA}
\altaffiltext{7}{Institute of Physics and Astronomy, University of Potsdam, 14476 Potsdam-Golm, Germany}
\altaffiltext{8}{DESY, Platanenallee 6, 15738 Zeuthen, Germany}
\altaffiltext{9}{Astronomy Department, Adler Planetarium and Astronomy Museum, Chicago, IL 60605, USA}
\altaffiltext{10}{School of Physics, National University of Ireland Galway, University Road, Galway, Ireland}
\altaffiltext{11}{Department of Astronomy and Astrophysics, 525 Davey Lab, Pennsylvania State University, University Park, PA 16802, USA}
\altaffiltext{12}{Department of Physics and Astronomy, Purdue University, West Lafayette, IN 47907, USA}
\altaffiltext{13}{Department of Physics and Astronomy, University of Utah, Salt Lake City, UT 84112, USA}
\altaffiltext{14}{School of Physics and Astronomy, University of Minnesota, Minneapolis, MN 55455, USA}
\altaffiltext{15}{Department of Physics, California State University - East Bay, Hayward, CA 94542, USA}
\altaffiltext{16}{Physics Department, McGill University, Montreal, QC H3A 2T8, Canada}
\altaffiltext{17}{Department of Physics and Astronomy and the Bartol Research Institute, University of Delaware, Newark, DE 19716, USA} 
\altaffiltext{18}{Physics Department, Columbia University, New York, NY 10027, USA}
\altaffiltext{19}{Santa Cruz Institute for Particle Physics and Department of Physics, University of California, Santa Cruz, CA 95064, USA}
\altaffiltext{20}{Department of Physics and Astronomy, University of Iowa, Van Allen Hall, Iowa City, IA 52242, USA}
\altaffiltext{21}{Department of Physics and Astronomy, DePauw University, Greencastle, IN 46135-0037, USA}
\altaffiltext{22}{School of Physics and Center for Relativistic Astrophysics, Georgia Institute of Technology, 837 State Street NW, Atlanta, GA 30332-0430}
\altaffiltext{23}{Department of Physics, Anderson University, 1100 East 5th Street, Anderson, IN 46012}
\altaffiltext{24}{Department of Physics and Astronomy, Barnard College, Columbia University, NY 10027, USA}
\altaffiltext{25}{Enrico Fermi Institute, University of Chicago, Chicago, IL 60637, USA}
\altaffiltext{26}{Department of Applied Science, Cork Institute of Technology, Bishopstown, Cork, Ireland}
\altaffiltext{27}{University of Maryland, College Park / NASA GSFC, College Park, MD 20742, USA}
\altaffiltext{28}{CIERA, Deprtment of Physics and Astronomy, Northwestern University, Evanston, IL 60208 USA}
\pagebreak

\begin{abstract}
The Galactic Center Ridge has been observed extensively in the past by both GeV and TeV gamma-ray instruments revealing a wealth of structure, including a diffuse component as well as the point sources G0.9+0.1 (a composite supernova remnant) and Sgr A* (believed to be associated with the supermassive black hole located at the center of our Galaxy). Previous very high energy (VHE) gamma-ray observations  with the H.E.S.S. experiment have also detected an extended TeV gamma-ray component along the Galactic plane in the $>$300 GeV gamma-ray regime.  Here we report on observations of the Galactic Center Ridge from 2010-2014 by the VERITAS telescope array in the $>$2 TeV energy range. From these observations we 1.) provide improved measurements of the differential energy spectrum for Sgr A* in the $>$2 TeV gamma-ray regime, 2.) provide a detection in the $>$2 TeV gamma-ray emission from the composite SNR G0.9+0.1 and an improved determination of its multi-TeV gamma-ray energy spectrum, 3.) report on the detection of VER J1746-289, a localized enhancement of $>$2 TeV gamma-ray emission along the Galactic plane. 
\end{abstract}

\section{Introduction}

The Galactic Center Ridge region, defined here as the central 3$^{\circ}$$\times$1$^{\circ}$ (\textit{l},\textit{b}) region of the Galactic plane, encompasses a large number of non-thermal sources which are known to accelerate particles to relativistic energies. The Galactic Center Ridge includes structures such as the non-thermal filaments (NTFs) seen in the radio band \citep{radioYSC,20cm,radiogbt,LaRosa,Nord}, multiple centers of hard X-ray emission \citep{Chandrasources, xraysc1, xraysc2, Goldwurm}, sources of MeV-GeV gamma-ray emission \citep{EGRET, 3FGL}, and both point-like and diffuse or extended structures seen in the TeV gamma-ray regime \citep{vanEldik}.

The bright radio source Sgr A* located at the Galactic Center is associated with a supermassive (4 $\times$ 10$^{6}$ M$_{\odot}$) black hole \citep{BHMASS, Ghez} and is known to be a luminous ($\sim$10$^{33}$ ergs/s) hard X-ray source, with transient outbursts up to $>$10$^{35}$ ergs/s. Evidence for energetic flares from the region have also been seen at GeV energies by $\textit{Fermi}$ \citep{FERMIATEL}; however, through a relatively long timeline of TeV gamma-ray observations (from 1995 to 2015) with imaging atmospheric Cherenkov telescopes (IACTs), the source is observed to be constant at TeV gamma-ray energies \citep{CANGAROO, Whipple, HESSGC, HESSSpec, MAGIC, MATTHIAS}. 

This apparent lack of variability from Sgr A* in the TeV gamma-ray regime has led to questions about whether the TeV source is actually associated with Sgr A* at all; alternative sources such as the supernova remnant (SNR) Sgr A East have been ruled out as being associated with TeV emission \citep{Acero}, while others, such as the pulsar wind nebula (PWN) G359.95-0.04 \citep{Wang}, are still possible alternatives \citep{Hinton}. It is also possible that, unlike active galactic nuclei (AGN), the dominant source of TeV gamma-ray emission comes from a wind termination or accretion shock \citep{AandD, Ball, Chern} associated with Sgr A*, but not directly with emission from a centrally located jet. While the issue of a firm identification of the central TeV source remains somewhat unclear, for the analysis presented in this work we will assume that the central TeV source is associated with Sgr A* and refer to it as such. We revisit this question in the discussion (Section 6).

\begin{figure*}[t]
\centering
\includegraphics[width=0.75\textwidth]{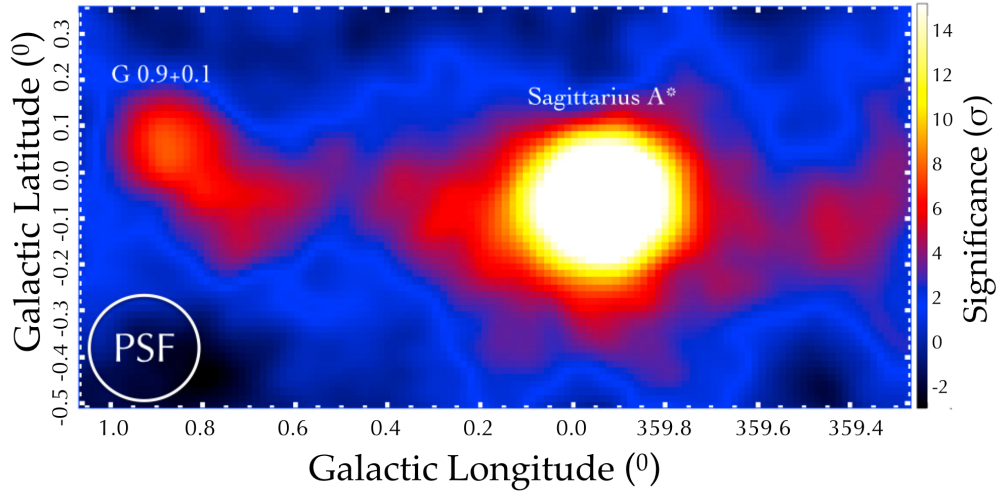}
\caption{ VERITAS $>$2 TeV gamma-ray significance map (smoothed with PSF) of the Galactic Center Ridge showing significant emission from Sgr A*, G0.9+0.1, and a diffuse emission region between these two point sources. The emission from the central region is saturated due to the color scale.}
\end{figure*}

\hypertarget{Figure1}{ } As well as providing measurements of the central TeV gamma-ray source (associated with Sgr A*), previous H.E.S.S. observations of the Galactic Center Ridge have also revealed significant TeV emission from the composite supernova remnant SNR G0.9+0.1 \citep{H.E.S.S.G0.9}, as well as two extended sources of TeV emission: HESS J1745-303 (associated with a  supernova remnant/molecular cloud interaction \citep{HESSSurvey} and HESS J1741-302 (unidentified, \citet{1741}). Additionally, the observation of a diffuse band of  $>$300 GeV gamma-ray emission centered on the Galactic Center by the H.E.S.S. collaboration \citep{HESSDiffuse}, which correlates well with the projected density of molecular clouds in the region, is strong evidence for the interaction between a population of cosmic rays accelerated near the Galactic Center and the material in the molecular clouds (presumably from hadronic interactions, see \citet{Egberts}). Under a cosmic-ray driven scenario,  the H.E.S.S. results can be explained by a density of comic rays 3-6 times higher near the Galactic Center than in the solar neighborhood; a reasonable conclusion is therefore that a source within the Galactic Center (possibly Sgr A*) has acted in the past as a source of Galactic cosmic rays which can then escape the Galactic Center Ridge and diffuse throughout the Galaxy.

In addition to these astrophysical processes, the Galactic Center is expected to be the densest local region of particle dark matter in our local universe. While difficult to disentangle from conventional astrophysical processes, dark matter annihilation could provide a measurable contribution to the radio through gamma-ray emission in a region extending several degrees from the Galactic Center. One of the suggested possibilities for particle dark matter is the lightest supersymmetric particle (the neutralino, $\chi_{0}$) which can, depending on the mass of the $\chi_{0}$ particle, self-annihilate into gamma rays with energies in the range of both GeV and TeV gamma-ray instruments. In depth searches for this annihilation signal in the TeV regime have been performed by IACTs in the Galactic Center region \citep{HESSGCDM, Abramowski2011, Aharonian2006}, strongly constraining the allowed thermally averaged cross-section for self-annihilation to values near the edge of the parameter space allowed by cosmological measurements and supersymmetry models. Within the last few years, evidence of a Galactic Center excess in the 1-3 GeV regime from $\textit{Fermi}$ observations has led to claims that this dark matter self-annihilation has already been detected from the Galactic Center \citep{Hooper1, MandK, Daylan}. While this prospect is tantalizing, it is of crucial importance to understand the various conventional, local sources which can contribute to both the GeV and TeV gamma-ray emission from the region. For example, the $\textit{Fermi}$-LAT measured excess could perhaps be explained by a population of millisecond pulsars near to the Galactic Center \citep{Weniger}. As the spectral shape of the $\textit{Fermi}$-LAT excess includes a cutoff below 100 GeV, this excess would not extend into the TeV gamma-ray regime; observations with instruments such as VERITAS and H.E.S.S. offer the ability to identify regions of the inner Galaxy which are shining in gamma rays from purely astrophysical (and not $\chi_{0}$ self-annihilation) processes. These regions can then be removed from future dark matter analyses with $\textit{Fermi}$-LAT.

In this work we present the VERITAS observations of the Galactic Center Ridge in the $>$2 TeV gamma-ray regime, taken from 2010 to 2014. From these observations, we report the measurement of the position and spectrum of the central source which is spatially coincident with Sgr A*, improving upon the earlier VERITAS measurements and H.E.S.S. measurement above 2 TeV. We also report the detection by VERITAS of the composite SNR G0.9+0.1 and the measurement of its spectrum in the  $>$2 TeV gamma-ray band, improving upon previous H.E.S.S. measurements in this energy range. Finally, we report the detection of a new source of $>$2 TeV gamma-ray emission near to the Galactic Center VER J1746-289.

\section{VERITAS Observations}
The Very Energetic Radiation Imaging Telescope Array System (VERITAS), located at the Fred Lawrence Whipple Observatory (FLWO) in southern Arizona (31$^{\circ}$ 40$^{\prime}$ N, 110$^{\circ}$ 57$^{\prime}$  W,  1.3 km above sea level) is an array of four 12-meter IACTs. Since the commissioning of the array in 2007, VERITAS has provided excellent angular resolution and sensitivity to TeV gamma-ray sources \citep{HolderVTS}. In normal operation (i.e. high elevation observations), VERITAS is sensitive in the energy range from 85 GeV to $>$30 TeV and is capable of detecting a 1$\%$ Crab nebula flux in approximately 25 hours of observation time. VERITAS has an energy resolution of 15$\%$ at 1 TeV and a typical angular resolution of $<$0.1$^{\circ} $.

Between 2010 and 2014, VERITAS accumulated $\sim$85 hours live time of quality-selected observations of the Sgr A* region. Due to the Northern Hemisphere location of VERITAS, the Sgr A* region never transits above 30$^{\circ}$ elevation. Therefore, this work has been performed using the ``Displacement" analysis method for TeV gamma-ray astronomy \citep{Whipple, MATTHIAS}, which utilizes the displacement between the center of gravity of a parameterized Hillas ellipse and the location of the shower position within the camera plane. This method compensates for the degradation in angular resolution (caused by small parallactic displacements between shower images) usually caused by observations taken by IACTs at small elevation angles. Through the use of the Displacement method, the VERITAS observations of Sgr A* have a point spread function of 0.12$^{\circ}$ (68$\%$ containment radius). Through Monte Carlo simulations we estimate an energy resolution of $\sim$25-30$\%$ for the observations detailed in this work.  Due to the Cherenkov light from incident gamma rays having to traverse a much larger atmospheric depth during large zenith angle observations, the light from lower energy showers is insufficient to trigger the array, resulting in an increased energy threshold of 2 TeV for the observations detailed in this work ($>$60$^{\circ}$ zenith angle). 

For the imaging analysis and significance calculations included in this work, the ring background model \citep{RBM} was used; for the spectral analysis the reflected-region model \citep{RRM} was employed. Both background estimation techniques were performed with masked regions corresponding to all known TeV sources within 4$^{\circ}$ of the Galactic Center: HESS J1747-281 (G0.9+0.1), HESS J1745-290 (Sgr A*), the extended sources HESS J1741-302 and HESS J1745-303, and the central diffuse TeV component lying along the Galactic plane.


These observations result in a significant detection of several distinct regions of $>$2 TeV gamma-ray emission in the Galactic Center Ridge (significance skymap shown in \hyperlink{Figure1}{Figure 1}). The brightest source within the field is the central source coincident with Sgr A*. These data also provide a strong detection of $>$2 TeV gamma-ray emission from a region corresponding to the composite supernova remnant G0.9+0.1, as well as an\hypertarget{Figure2}{ } extended  component of emission along the Galactic plane. In sections 3-5 we examine these detections and provide skymaps and spectra for both Sgr A* and G0.9+0.1, and we present the detection of VER J1746-289: a new VERITAS source of TeV gamma rays embedded within the extended component along the plane.

\begin{figure*}[t]
\centering
\includegraphics[width=1.0\textwidth]{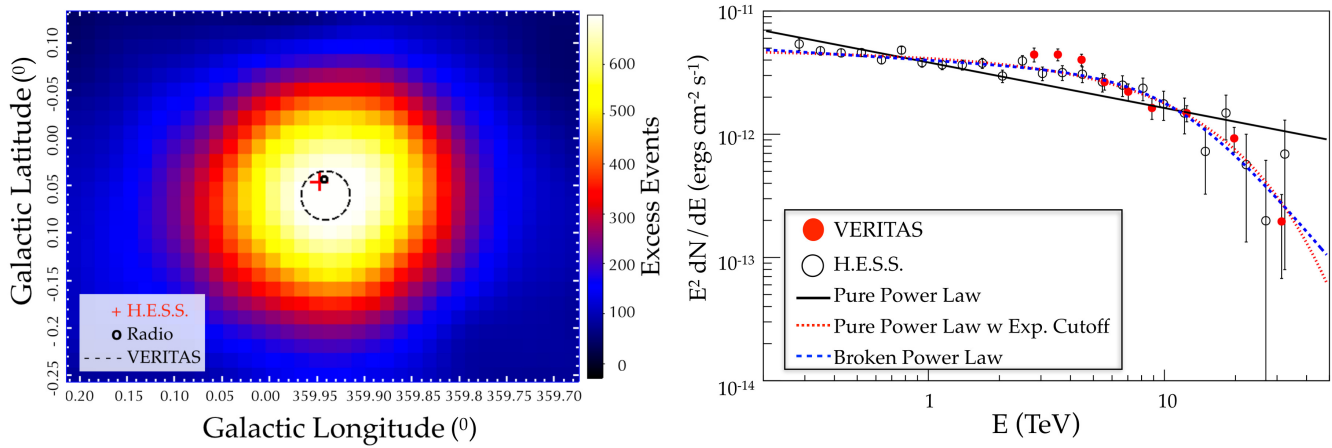}
\caption{The VERITAS $>$ 2 TeV gamma-ray excess map of Sgr A* showing the VERITAS source location compared to the H.E.S.S. and radio locations (left). The black dashed circle represents the total error on the VERITAS fit. On the right is shown the differential energy spectrum using both VERITAS and H.E.S.S. points along with the model fits described in the text.}
\label{fullmap2}
\end{figure*}

\begin{table*}[t]
\small
 \centering
  \caption{The results of the fitting of the VERITAS and H.E.S.S. spectral point of Sgr A* described in the text. }
 \begin{tabular}{c|c|c|c|c|c}
\hline
\textbf{\hypertarget{Table1}{Model}}&  \textbf{N$_{0}$ (cm$^{-2}$ s$^{-1}$ TeV$^{-1}$)} & \textbf{$\Gamma_{1}$} & \textbf{$\Gamma_{2}$} & \textbf{E$_{break}$ or E$_{cut}$ (TeV)} & \textbf{$\frac{\chi^{2}}{n.d.f.}$}   \\\hhline{=|=|=|=|=|=}
Power Law & 2.36 ($\pm$ 0.05) $\times$ 10$^{-12}$ & 2.37 $\pm$ 0.02 & N/A & N/A &148/32\\\hline
Exp. Cutoff  & 2.82 ($\pm$ 0.08) $\times$ 10$^{-12}$ & 2.05 $\pm$ 0.04 & N/A & 12.1 $\pm$ 1.6 & 35/31 \\
Power Law &  && &  & \\\hline
Smoothly Broken & 2.55 ($\pm$ 0.07) $\times$ 10$^{-12}$ &2.14 $\pm$ 0.04 & 4.39 $\pm$ 0.39 & 12.1 $\pm$ 1.7 & 32/30 \\
Power Law & &&  &  \\\hline 

\end{tabular}

\end{table*}
\normalsize

\section{VER J1745-290 (Sgr A*)}
In an analysis of earlier VERITAS observations \citep{MATTHIAS}, VER J1745-290 (coincident with Sgr A*) was detected at a statistical significance of 18 standard deviations (18 $\sigma$) in approximately 46 hours of observations between 2010 and 2012. In the total of 85 hours of observations reported in this work, VERITAS detected a total of 735 excess gamma-ray events from VER J1745-290, resulting in a detection significance of $>$25$\sigma$. The resulting $>$2 TeV gamma-ray excess map is shown in \hyperlink{Figure2}{Figure 2 (left)} along with both the H.E.S.S. ($>$300 GeV) \citep{Acero} and radio locations \citep{radiolocation} of Sgr A*. The refined VERITAS position of VER J1745-290 is l = 359.94$^{\circ}$ $\pm$ 0.002$^{\circ}_{stat}$ $\pm$ 0.013$^{\circ}_{sys}$, b = -0.053$^{\circ}$ $\pm$ 0.002$^{\circ}_{stat}$ $\pm$ 0.013$^{\circ}_{sys}$, in good agreement with both the radio and H.E.S.S. positions. 

The VERITAS differential energy spectrum of Sgr A* from 2 to 30 TeV (derived from an integration region of 0.13$^{\circ}$ centered on VER J1745-290) is shown, along with the H.E.S.S. spectral points \citep{HESSSpec} in \hyperlink{Figure2}{Figure 2 (right)}. While the H.E.S.S. observations allow for very rich statistics at lower energies, the large effective area for large zenith angle VERITAS observations of Sgr A* provide a significant improvement in statistics at multi-TeV energies. By providing a joint fit to both the H.E.S.S. and VERITAS points from 0.2 to 50 TeV, more refined spectral model parameters can be obtained. Following the analysis of \citet{HESSSpec}, we investigated spectra following the shape of 1.) a pure power law, 2.) a power law with an exponential cutoff, and 3.) a smoothly broken power law. These functions have forms of (respectively): 

\begin{figure*}[t]
\centering
\includegraphics[width=1.0\textwidth]{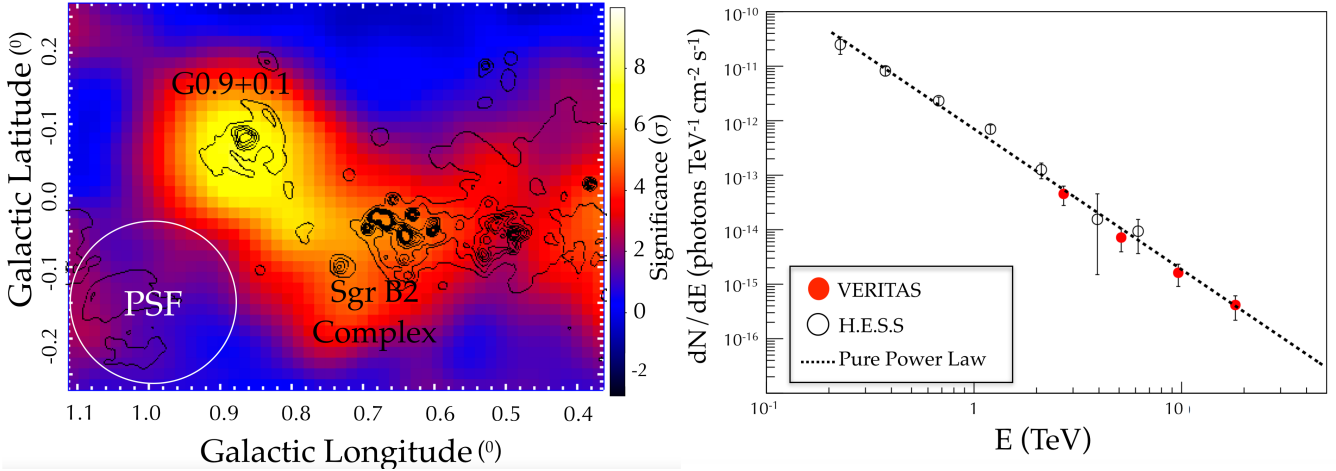}
\caption{Left: The VERITAS $>$2 TeV gamma-ray significance map (smoothed with PSF) of the composite SNR G0.9 +0.1, along with VLA 20cm radio contours \citep{20cm}. The VERITAS location is consistent with the core of the composite SNR. Excess TeV gamma-ray emission from the Sgr B2 region can also be seen adjacent to G0.9+0.1. Right: The differential energy spectrum of G0.9+0.1 using both VERITAS and H.E.S.S. \citep{H.E.S.S.G0.9} data points. The spectrum is well fit by a simple power law with no indication of a cutoff up to $\sim$20 TeV.}
\label{fullmap3}
\end{figure*}

\small
\begin{equation}
\frac{dN}{dE} = N_{0} \times \big(\frac{E}{1 \hspace{0.5mm}\mathrm{TeV}}\big)^{-\Gamma_{1}}
\end{equation}

\begin{equation}
\frac{dN}{dE} = N_{0} \times \big(\frac{E}{1 \hspace{0.5mm}\mathrm{TeV}}\big)^{-\Gamma_{1}} \times e^{\frac{-E}{E_{cut}}}
\end{equation}

\begin{equation}
\frac{dN}{dE} = N_{0} \times \big(\frac{E}{1 \hspace{0.5mm}\mathrm{TeV}}\big)^{-\Gamma_{1}} \times \frac{1}{1+(\frac{E}{E_{break}})^{\Gamma_{2}-\Gamma_{1}}}
\end{equation}
\normalsize
The fitting results (shown in \hyperlink{Table1}{Table 1}, and in \hyperlink{Figure2}{Figure 2, right}) clearly disfavor a pure power-law fit, while the exponential-cutoff power law and smoothly broken power law models both provide adequate fits (reduced $\chi^{2}$ values close to 1.0) with similar values for cutoff/break energies. In the case of a power law with an exponential cut-off, the spectral parameters are in good agreement with \cite{HESSSpec} and refine the location of the spectral cutoff (E$_{cutoff}$) to 12.1 $\pm$ 1.6 TeV. It is important to note that the measured spectrum includes contributions from any sources that may fall within 0.13$^{\circ}$ of Sgr A* (such as a diffuse component). Here we conservatively estimate the systematic error to be approximately 40$\%$ on the energy scale and 40$\%$ on the flux normalization (N$_{0}$) (see \citealt{MATTHIAS} for a description of how this error is derived). Taken in quadrature with the associated H.E.S.S. systematic errors reported in \citep{HESSSpec}, the VERITAS systematic errors dominate, resulting in a total estimated systematic error for the above fits of $\sim$40$\%$ on energy scale, and $\sim$40$\%$ on flux normalization.

Assuming that either an exponential-cutoff power law or broken power law provides the best model of the spectrum from the source, the determination of the cutoff/break energy in the TeV gamma-ray spectrum provides an important mechanism to study particle acceleration in the region. We note that the overlap between multiple components of TeV gamma-ray emission in the area (Sgr A*, diffuse, and possibly others) makes a simple extraction of the diffuse component from the Sgr A* spectrum problematic.  As such, we take the approach of only using the observed TeV gamma-ray spectrum from the direction of Sgr A* to compare to emission models in the discussion section at the end of this work. We note that even with this analysis choice,  our current result is consistent with the results of \citet{Viana} in which the authors perform a subtraction of a modeled diffuse component from the Sgr A* TeV gamma-ray spectrum and find a spectral cutoff of 10.7 $\pm$ 2.0$_{stat}$ TeV.

\section{VER J1747-281 (G0.9+0.1)}

The composite supernova remnant G0.9+0.1 consists of a bright, compact radio PWN core surrounded by an extended radio shell \citep{g09radio} and is estimated to have an age of a few thousand years \citep{Mezger, H.E.S.S.G0.9}. G0.9+0.1 was first announced as a TeV gamma-ray source by the H.E.S.S. collaboration \citep{H.E.S.S.G0.9}, detecting $>$200 GeV gamma-ray emission at the level of approximately 2$\%$ of the Crab nebula flux. The H.E.S.S. source is attributed to the core of the remnant due to the observed morphology, as well as the apparent lack of strong hard X-ray emission in the shell remnant. The H.E.S.S. spectrum of the source from 0.2 to 7 TeV is well fit by a simple power law with a spectral index of 2.4 $\pm$ 0.11$_{stat}$.

VERITAS observations of the Galactic Center Ridge taken during 2010-2014 also allow for a statistically significant detection of G0.9+0.1 in the $>$2 TeV gamma-ray regime. In the 85 hours of observations reported in this work, VERITAS detected a total of 134 excess events from G0.9+0.1, corresponding to a statistically significant detection at the 7$\sigma$ level. The VERITAS source position (\hyperlink{Figure3}{Figure 3, left}) is centered at l = 0.86$^{\circ}$ $\pm$ 0.015$^{\circ}_{stat}$ $\pm$ 0.013$^{\circ}_{sys}$, b = 0.067$^{\circ}$ $\pm$ 0.02$^{\circ}_{stat}$ $\pm$ 0.013$^{\circ}_{sys}$ and is given the VERITAS source name VER J1747-281. The VERITAS position is coincident with both H.E.S.S. position and the radio core location. The joint VERITAS and H.E.S.S. spectra of G0.9+0.1 (\hyperlink{Figure3}{Figure 3, right}) from 0.2 to 30 TeV are well fit (reduced $\chi^{2}$ of 3.1/9) by a pure power law (Eq. 3.1) with normalization (at 1 TeV) of 7.07 $\pm$ 0.66$_{stat}$ $\times$ 10$^{-13}$ photons TeV$^{-1}$ cm$^{-2}$ s$^{-1}$ and index of 2.51 $\pm$ 0.07$_{stat}$, consistent with the H.E.S.S. measurement alone. Adding the systematic errors for both the H.E.S.S. and VERITAS measurements in quadrature, we arrive at an error of 41$\%$ on the spectral index and 45$\%$ on the flux normalization. We find no strong indications of a spectral break up to $\sim$20 TeV. 


\hypertarget{Figure3}{ }

\section{VER J1746-289}

In \citet{HESSDiffuse}, the H.E.S.S. collaboration presented residual maps (i.e. after subtracting known point sources within the field of view) of the $>$300 GeV gamma-ray emission from the Galactic plane. These residual maps revealed a complicated network of diffuse gamma-ray emission within the central 3$^{\circ}$ of the plane. When plotted along with the CS emission contours \citep{NANTEN}, the H.E.S.S. emission appears correlated with dense molecular cloud regions (bright in CS line emission). \hypertarget{Figure4}{} However, given the complicated nature of the region, this measurement was unable to rule out the possibility of a significant contribution to the TeV gamma-ray flux coming from unresolved point sources. 

\begin{figure}[t]
  \begin{center}
    \includegraphics[width=0.5\textwidth]{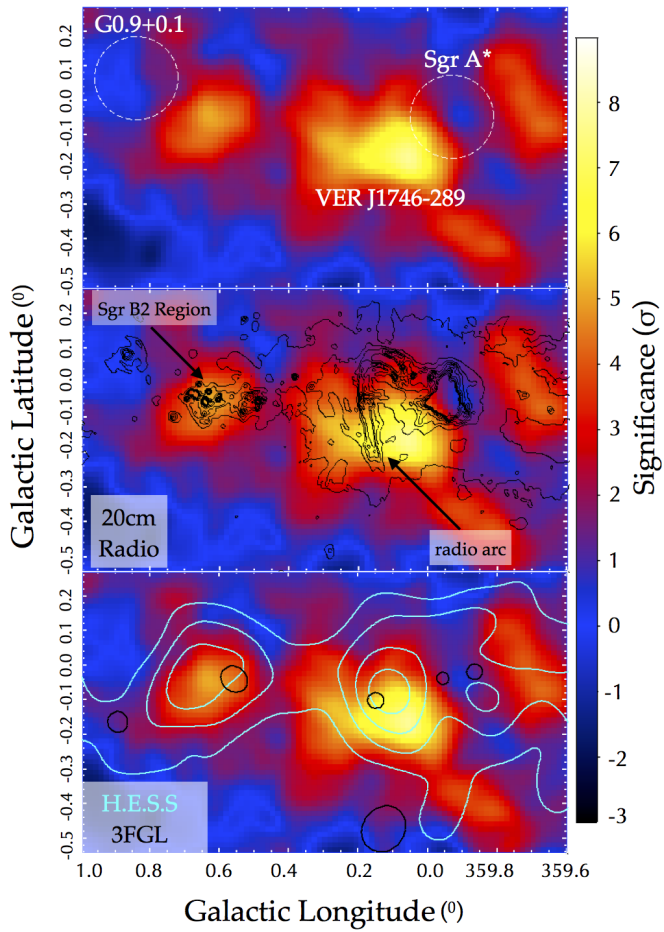}
  \end{center}
  \caption{The VERITAS $>$2 TeV gamma-ray significance maps (smoothed with PSF) of the Galactic Center Ridge after subtracting excess emission from Sgr A* and G0.9+0.1. The top panel shows the locations of the subtracted point sources as well as the VERITAS source VER J1746-289. VLA 20cm radio contours from \citet{20cm} are shown in the middle panel, with H.E.S.S. 275, 300, 325, and 350 excess event contours \citep{HESSDiffuse} and \textit{Fermi}-LAT 3FGL \citep{3FGL} sources shown in the bottom panel.}
\end{figure}

To investigate whether the H.E.S.S. residual component is present in the $>$2 TeV gamma-ray VERITAS skymaps, we effectively masked the two point sources (Sgr A* and G0.9 +0.1) by removing the excess counts from a 0.12$^{\circ}$ region surrounding their best fit locations. The resulting significance skymap is shown in  \hyperlink{Figure4}{Figure 4} with radio (middle panel), and \textit{Fermi}-LAT/H.E.S.S. intensity contours (bottom panel) overlaid. A band of  $>$2 TeV gamma-ray emission reaches $\sim$1$^{\circ}$ to the east of Sgr A*. This morphology is consistent with the result of \cite{HESSDiffuse} for emission above 300 GeV, with two main enhancements: the first co-located with the giant molecular cloud complex Sgr B2 (see \hyperlink{Figure3}{Figure 3, left}; \hyperlink{Figure4}{Figure 4}) and a second region directly adjacent to Sgr A*. With respect to the enhancement seen near Sgr B2, the VERITAS data indicates a $>$2 TeV gamma-ray excess at a statistical significance of 5.3$\sigma$. After an appropriate trials factor to account for analysis cuts (signal versus noise selection criteria) and a PSF-sized search region are applied, this significance decreases to 4.1$\sigma$, bringing the excess  below the threshold for a claim of an individual source detection (5$\sigma$) by VERITAS.

In this work we choose to focus on the localized excess of $>$2 TeV gamma-ray emission within this diffuse component, directly adjacent to Sgr A*. The center of this excess is located at l = 0.055$^{\circ}$ $\pm$ 0.01$^{\circ}_{stat}$ $\pm$ 0.013$^{\circ}_{sys}$, b = -0.148$^{\circ}$ $\pm$ 0.01$^{\circ}_{stat}$ $\pm$ 0.013$^{\circ}_{sys}$ and is given the name VER J1746-289. VER J1746-289 is well fit (reduced $\chi^{2}$ of 216/172) by an asymmetric two-dimensional Gaussian of $\sigma_{l}$=0.08$^{\circ}$, $\sigma_{b}$=0.03$^{\circ}$ (rotation angle of -15.4$^{\circ}$ with respect to Galactic Latitude), therefore the source is only marginally extended in Galactic Longitude. VER J1746-289 is detected at a statistical significance of 7.6$\sigma$ before applying a trials factor. Using an appropriate search region (and accounting for analysis cuts) this significance reduces to 6.7$\sigma$. The proximity of VER J 1746-289 to the bright excess of Sgr A* might naturally cause a concern that VER J1746-289 might be an artifact of the source subtraction procedure utilized in this work. To address this concern, the size of the subtraction region used  for the residual maps was modified by $\pm$ 20$\%$ with no significant difference caused in the resulting morphology of VER J1746-289. Additionally, we note that the same source subtraction procedure was used for G0.9+0.1 with no residual features created (see \hyperlink{Figure4}{(Figure 4, top panel)}).

\section{Discussion}

\subsection{Point Sources in the Galactic Center Ridge} In this work we have presented the observations made by VERITAS revealing the complex morphology of the Galactic Center Ridge region at multi-TeV gamma-ray energies. Using large zenith angle observations, we obtain excellent sensitivity above 2 TeV, complementing the lower energy threshold measurements made by the H.E.S.S. collaboration. The VERITAS detections of both Sgr A* and G0.9+0.1 above 2 TeV, in conjunction with the previous H.E.S.S. observations of these regions allow for more statistically rich spectral measurements of these sources, providing better constraints on modeling the emission processes at work in both Sgr A* and G0.9+0.1.

 In the case of G0.9+0.1, the lack of a break in the spectrum up to $\sim$20 TeV as well as the lack of emission seen by \textit{Fermi}-LAT), in conjunction with X-ray measurements, can be used to help  constrain models of emission in this PWN.

\begin{figure*}[t]
\centering
\includegraphics[width=0.65\textwidth]{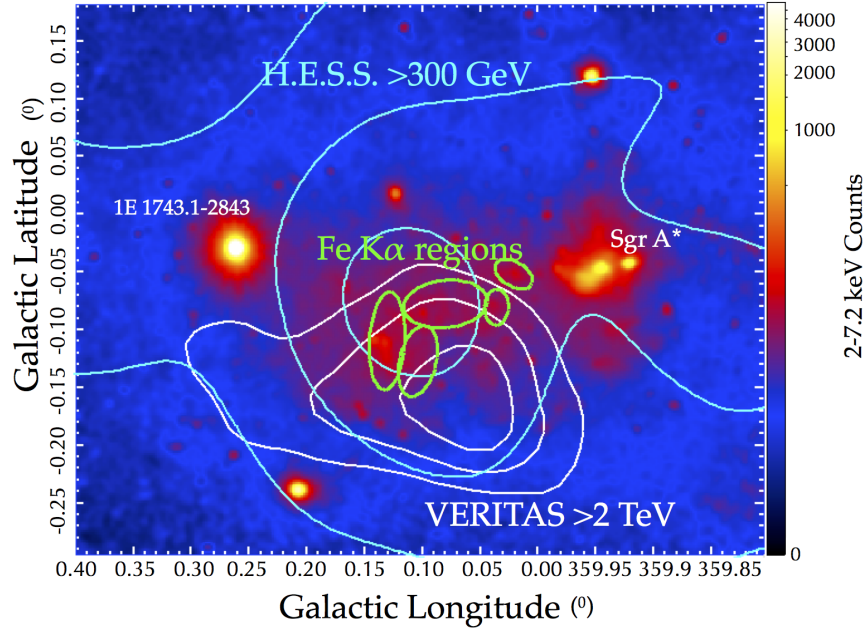}
\caption{The XMM-Newton hard X-ray (2-7.2 keV) map of the region adjacent to Sgr A*. The H.E.S.S. excess event contours (cyan) are shown along with the VERITAS 5,6,7 $\sigma$ significance contours (white) for the source subtracted maps shown in Figure 4. Also shown (green ellipses) are regions identified by Fe K emission to be particularly dense molecular cloud regions lit up by a superluminal shock.}
\label{fullmap}
\end{figure*}
The TeV gamma-ray emission from Sgr A* still lacks definitive explanation. The measurement of the spectrum of Sgr A* by both VERITAS and H.E.S.S. at $\sim$10 TeV implies a primary spectrum of particles with energies beyond 100 TeV \citep{HESSSpec}, the generation of which can be accommodated within a range of both hadronic \citep{Ball, Chern} and leptonic \citep{AandD} scenarios (see \citet{MATTHIAS} for further examination of these models as compared to the observed broadband spectral energy distribution).

 
It is clear that Sgr A* is not as simply modeled as other point sources of TeV gamma-ray emission within the Galaxy. For instance, it is estimated that Sgr A* is relatively underluminous for its estimated mass ($\sim$8 orders of magnitude below its Eddington luminosity). This has led many to question whether this present (relatively quiescent) state has been punctuated in the past by outbursts from the central engine. Observations of the $\textit{Fermi}$-LAT GeV haze symmetrically mirrored above and below Sgr A* (the so-called ``Fermi bubbles", see \citet{Bubble2} and \citet{Bubble1}) are the most direct evidence for previous increased activity from Sgr A*. Additional evidence for previous transient outbursts from Sgr A* includes X-ray observations which indicate that the nearby ($\sim$100 pc from Sgr A*) giant molecular cloud complex Sgr B2 is the echo site of a previous flare event from Sgr A* around 300 years ago \citep{XRAYSGRB21,XRAYSGRB22}, suggesting that Sgr A* may recently have been more luminous than its current level by four orders of magnitude. More recently (and closer to Sgr A*), XMM-Newton observations indicate that molecular clouds nearby to Sgr A* have been irradiated by a transient event sometime within the last 100 years \citep{Ponti}.  Along these lines, it is natural to continue to utilize VHE observations to search for variability in the gamma-ray band: a detection of TeV variability by itself or correlated with X-ray flares can help to discriminate among different emission models \citep{MATTHIAS}, as well as provide crucial links to this poorly understood transient behavior in other wavelengths. 


\subsection{Residual Emission}
We have also presented $\textit{source subtracted}$ residual maps of the Galactic Center Ridge, removing emission components from both G0.9+0.1 and Sgr A* in order to reveal less prominent TeV features along the Galactic plane (\hyperlink{Figure4}{Figures 4 }and \hyperlink{Figure5}{5}).  This has revealed structure above several TeV extending along the plane to the east of Sgr A*, thus confirming and extending in energy the H.E.S.S. result. One of the local enhancements within this extended structure, VER J1746-289,  is significantly detected by VERITAS and is most likely associated with known non-thermal structures (radio, X-ray, GeV, TeV) in the region.  

VER J1746-289 is coincident with multiple regions of radio - TeV emission, including the well-studied Galactic radio arc \hyperlink{Figure4}{(Figure 4, middle panel)}, as well as the \textit{Fermi}-LAT 3FGL source \hyperlink{Figure4}{(Figure 4, bottom panel)} 3FGL J1746.3-2851c. The morphology of VER J1746-289 is also consistent with the H.E.S.S. excess in the region \hyperlink{Figure4}{(Figure 4, bottom panel}; \hyperlink{Figure5}{Figure 5)}. However, we note that the peak significance of  VER J1746-289 appears to be offset from the H.E.S.S. peak excess. As the VERITAS and H.E.S.S. excesses are derived from differing energy regimes ($>$2 TeV vs $>$300 GeV), it is possible that some variation in morphology might appear when comparing the two maps. In \citet{Lemiere}, the H.E.S.S. collaboration presented a 2D likelihood analysis of over 250 hours of observations which indicates that the peak excess seen in their maps is due to a single point source (possibly correlated with a pulsar wind nebula), and not due to the structure of the diffuse emission surrounding the Galactic Center. In light of this new result, it is likely that the structure of VER J1746-289 is due to a superposition of emission from this new H.E.S.S. point source and a local enhancement in the diffuse emission near the Galactic Center.  
 
The similarity in morphology of the $>$300 GeV \citep{HESSDiffuse} and $>$2 TeV gamma-ray skymaps is consistent with the previous interpretation of the H.E.S.S. emission as coming from $\pi^{0}$ emission from a hard spectrum cosmic ray population. To argue for this hypothesis, an analysis of the H.E.S.S. Galactic plane survey data \citep{Egberts} shows that the diffuse TeV gamma-ray emission along the plane can be well explained by cosmic ray protons interacting with HI and HII regions, resulting in gamma rays through neutral pion decay. In the analysis of \citet{Egberts} there is only a small secondary contribution from unresolved point sources and inverse-Compton scattering by leptonic cosmic rays. 

In addition to protons accelerated at or near Sgr A*, another possibility to explain the origin of the non-thermal emission in the Galactic Center Ridge is the interaction of synchrotron emitting electrons with molecular gas \citep{Pohl1, Farhad2013}. As opposed to the hadronic scenario, ample evidence for the presence of relativistic electrons already exists: There is a strong source of diffuse synchrotron emission from the central few hundred parsecs of the Galaxy \citep{LaRosa,Pohl2}, as well as a population of non-thermal radio filaments \citep{Farhad2013, 20cm}. There is also a vast amount of diffuse H$_3^+$ distributed in the Galactic center region, implying cosmic ray ionization rate one to two orders of magnitude higher in the Galactic center than in the Galactic disk \citep{Oka, LePetit}.



It is clear that transient outbursts from Sgr A* must also play a role in the overall non-thermal emission from the region within any (leptonic or hadronic) model. Flickering 6.4 keV Fe K$\alpha$ emission has been observed from the dense molecular cloud regions near Sgr A* as well as further out near Sgr B2 \citep{Ponti, XRAYSGRB21,XRAYSGRB22}. This observation provides evidence for transient activity producing significant amounts of energetic particles, this behavior exhibiting variability over a timescale of several years. In \hyperlink{Figure5}{Figure 5} we show the VERITAS (significance) and H.E.S.S. (excess) contours of the region overlaid on the XMM-Newton 2-7.2 keV count map. The known, non-thermal X-ray filamentary structure is clearly apparent, that overlaps well with the GeV/TeV gamma-ray emission in the region. \hyperlink{Figure5}{Figure 5} also shows the regions in this map which exhibit strong Fe K$\alpha$ emission lines \citep{Ponti}, corresponding to dense molecular cloud regions. In \citet{Ponti}, it was shown that the X-ray emission from this region includes a superluminal light front traveling through a molecular cloud region, powered by a local source (e.g. Sgr A*, or another nearby energetic source).

Despite the spatial overlap between the TeV gamma-ray emission and the variable X-ray regions, shown in \hyperlink{Figure5}{Figure 5}, it is difficult to produce a unified physical description that would explain both features. However, as Fe K$\alpha$ emission results from the ionization of neutral iron by energetic particles, measurements in this band provide a direct tracer for both high-density molecular cloud formations as well as the path of energetic electron transport on several year timescales. Thus, monitoring of this complex region across the non-thermal spectrum with radio to TeV instruments might reveal further variable behavior that can clarify the overall understanding of this complicated region. Furthermore, as additional data with VERITAS is accrued, the morphology of VER J1746-289 can be studied in greater detail. This will allow the determination of whether this source of TeV emission is primarily due to a point source (PWN) immersed in diffuse emission (as H.E.S.S. observations would suggest) or originating from the same dense molecular regions traced out by Fe K$\alpha$ and continuum X-ray emission. In the latter case, VER J1746-289 could prove to be a novel laboratory to study the high energy particle acceleration at very close distances to the Galactic Center.

\section{Acknowledgments}
This research is supported by grants from the U.S. Department of Energy Office of Science, the U.S. National Science Foundation and the Smithsonian Institution, and by NSERC in Canada. We acknowledge the excellent work of the technical support staff at the Fred Lawrence Whipple Observatory and at the collaborating institutions in the construction and operation of the instrument.

 A.W. Smith acknowledges support through the Cycle 7 \textit{Fermi} Guest Investigator program, grant number NNH13ZDA001N. 

The VERITAS Collaboration is grateful to Trevor Weekes for his seminal contributions and leadership in the field of VHE gamma-ray astrophysics, which made this study possible.

\bibliography{refs}

\end{document}